\documentclass{iaus}
\usepackage{graphicx}

\title[Eclipsing binary stars] %% give here short title %%
{Eclipsing Binary Stars as Tests of Stellar Evolutionary Models 
and Stellar Ages}

\author[Stassun et al.]   %% give here short author list %%
{Keivan G.\ Stassun$^1$,
Leslie Hebb$^2$, Mercedes L\'{o}pez-Morales$^3$, \and Andrej Pr\v{s}a$^4$}

\affiliation{$^1$Physics \& Astronomy Dept., Vanderbilt University,
VU Station B 1807, Nashville, TN 37235 \\ email: {\tt keivan.stassun@vanderbilt.edu} 
\\[\affilskip]
$^2$University of St.\ Andrews, $^3$Carnegie Institution of Washington,
$^4$Villanova University}

\pubyear{2008}
\volume{258}  %% insert here IAU Symposium No.
\jname{Ages of Stars}
\editors{E.E. Mamajek \& D. Soderblom, eds.}
\begin{document}

\maketitle

\begin{abstract}
Eclipsing binary stars provide highly accurate measurements 
of the fundamental physical properties of stars. They therefore serve 
as stringent tests of the predictions of evolutionary models
upon which most stellar age determinations are based. Models generally
perform very well in predicting coeval ages for eclipsing binaries
with main-sequence components more massive than $\approx 1.2$ M$_\odot$; 
relative ages are good to $\sim 5$\% or better in this mass regime. 
Low-mass main-sequence stars ($M < 0.8$ M$_\odot$) reveal large discrepancies 
in the model predicted ages, primarily due to magnetic activity in the 
observed stars that appears to inhibit convection and likely causes the radii 
to be 10--20\% larger than predicted. In mass-radius diagrams these stars 
thus appear 50--90\% older or younger than they really are. 
Aside from these activity-related effects, low-mass 
pre--main-sequence stars at ages $\sim 1$ Myr can also show non-coevality of 
$\sim 30$\% due to star formation effects, however these effects are
largely erased after $\sim 10$ Myr. 
\keywords{stars: binaries: eclipsing, stars: fundamental parameters, 
stars: activity}
\end{abstract}

\firstsection % if your document starts with a section,
\section{Introduction}
Eclipsing binary stars are one of nature's best laboratories for
determining the fundamental physical properties of stars and thus for
testing the predictions of theoretical models.
Detached, double-lined eclipsing binaries (hereafter EBs)
%\footnote{We will not discuss contact EBs (e.g.\ W UMa systems)
%or other complex or evolved systems.}) 
yield direct and accurate measures of the masses, radii, surface gravities,
temperatures, and luminosities of the two stars. These are measured
directly via combined analysis of multi-band light curves and
radial velocity measurements (\cite[Wilson \& Devinney 1971]{wilson71};
\cite[Pr\v{s}a \& Zwitter 2005]{prsa05}).
%required---in fact, the distance can be {\it measured} by comparing the
%measured luminosity with the observed apparent brightness.

\begin{figure}[ht]
% \vspace*{-2.0 cm}
\begin{center}
 \includegraphics[width=5.3in]{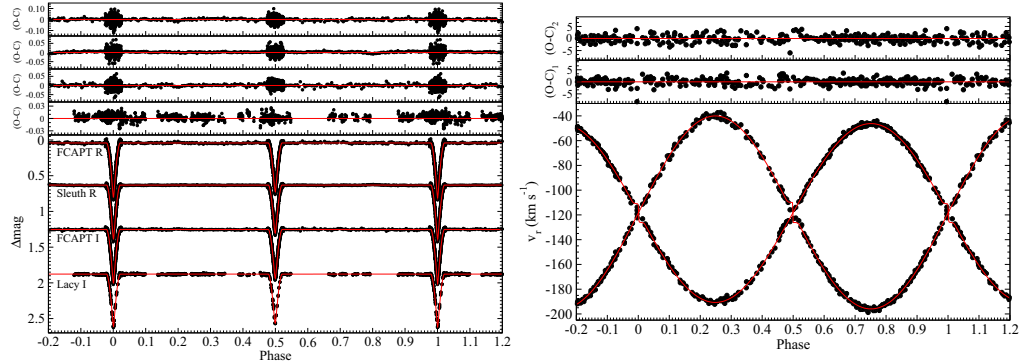} 
% \vspace*{-1.0 cm}
 \caption{Simultaneous analysis of multi-band light curves (left) and radial
velocities (right) of CM~Dra by \cite[Morales et al.\ (2008)]{morales08}. 
The resulting masses and radii of the stellar components are determined 
with an accuracy better than 0.5\%.}
   \label{CMDra}
\end{center}
\end{figure}

Knowledge of the distance to the EB is not required, and
thus the physical properties of the stars 
can be measured with exquisite accuracy. As an example, 
\cite[Morales et al.\ (2008)]{morales08} have measured the component 
masses and radii of the low-mass EB CM~Dra (Fig.\ \ref{CMDra}) with 
an accuracy better than 0.5\%, perhaps the most accurate 
measurements ever made for a low-mass EB. Similar accuracy has been
achieved for the high mass $\beta$ Aur 
(\cite[Southworth et al.\ 2007]{southworth07}).
Indeed, at this level of precision, 
a non-negligible contributor to the error budget is the uncertainty on 
Newton's gravitational constant (\cite[Torres \& Ribas 2002]{torres02}).

Such high quality measurements allow
the predictions of theoretical models to be rigorously tested.  For a
main-sequence star of a given mass and metallicity, the radius is a monotonic
function of age. Thus the models should assign the same age to the components
of an EB (i.e., they should lie on a single model isochrone in,
e.g., the $M$--$R$ plane), assuming that the components formed from the same
material at the same time. The apparent difference in age, $\Delta\tau$,
of the two components is thus a direct measure of the error
in the age calibration of the models. As we now discuss, the accuracy of
the age calibration is principally a function of stellar mass, varying
from $\sim 5$\% for $M>1.2$ M$_\odot$, to $\sim 10$\% for $M\approx 1$
M$_\odot$, to 50--90\% for $M<0.8$ M$_\odot$ (see Table~\ref{tab1}).
 
\begin{table}[b]
  \begin{center}
  \caption{Accuracy of stellar age calibrations from eclipsing binaries.}
  \label{tab1}
 {\scriptsize
  \begin{tabular}{|c|c|c|c|c|}\hline 
{\bf Regime} & {\bf Accuracy} & {\bf Limiting Physics and/or Data} & {\bf Exemplar(s)} & {\bf Refs.} \\ 
  &  $\Delta\tau^\dagger$ & &  &  \\ \hline
$M > 1.2\; {\rm M}_\odot$ & $\sim 5$\% & modeling of core overshooting, mixing & $\xi$ Phe & 1 \\ \hline
$\sim 1\; {\rm M}_\odot$ & $\sim 10$\% & modeling of convection, activity & CV Boo & 2 \\ 
   &   &  & V1174 Ori & 3 \\ \hline
$< 0.8\; {\rm M}_\odot$ & 50--90\% & activity calibration, abundance measurements & YY Gem& 4 \\ 
   &   &  & V818 Tau & 5 \\ 
   &   &  & V1061 Cyg & 6 \\ \hline
PMS stars, $\tau < 10$ Myr$^{\dagger\dagger}$ & $\sim 50$\% & star formation effects (e.g.\ accretion history) & Par 1802 & 7 \\ \hline
  \end{tabular}
  }
 \end{center}
\vspace{1mm}
 \scriptsize{
 {\it Notes:}\\
  $^\dagger$Apparent age difference of presumably coeval stellar components. 
  $^{\dagger\dagger}$Convolved with mass-dependent effects. \\
  $^1$\cite[Young \& Arnett (2005)]{young05},
  $^2$\cite[Torres et al.\ (2008)]{torres08},
  $^3$\cite[Stassun et al.\ (2004)]{stassun04},
  $^{\rm 4,5}$\cite[Torres \& Ribas (2002)]{torres02},
  $^6$\cite[Torres et al.\ (2006)]{torres06},
  $^7$\cite[Stassun et al.\ (2008)]{stassun08}
 }
\end{table}

\section{Accuracy of the stellar age calibration as a function of stellar
mass}

\subsection{Massive stars ($M > 1.2$ M$_\odot$)}

In general, theoretical models perform best in predicting coeval ages
in main-sequence EBs with $M > 1.2\; {\rm M}_\odot$.
For example, \cite[Young \& Arnett (2005)]{young05} have performed a
comprehensive re-analysis of the 20 EBs with $22 < M/{\rm M}_\odot
< 1.2$ that were included in the seminal review of \cite[Andersen
(1991)]{andersen91}. Their {\sc tycho} models incorporate updated abundances
and, most importantly, improved treatment of interior mixing physics such
as core convective overshooting. They find $\Delta\tau < 5$\% for the
typical case and $\Delta\tau < 10$\% for all of the EBs in their sample.

\begin{figure}[t]
% \vspace*{-2.0 cm}
\begin{center}
 \includegraphics[width=3.75in]{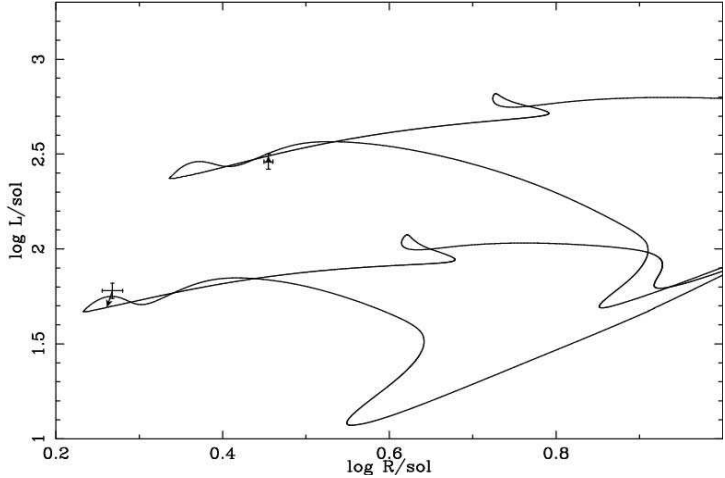} 
% \vspace*{-1.0 cm}
 \caption{Theoretical evolutionary model fits to the components of the
EB $\xi$~Phe by \cite[Young \& Arnett (2005)]{young05}. This massive,
slightly evolved EB is a challenging case, yet even without {\it ad hoc}
adjustments to the fitting parameters achieves $\Delta\tau\approx 3$\%.
 }
   \label{XiPhe}
\end{center}
\end{figure}

This excellent performance of the models includes a few EBs near the
terminal-age main sequence (TAMS), where the stars are evolving very rapidly
toward the red giant phase and for which any discrepancies in the models
are amplified. For example, Fig.\ \ref{XiPhe} shows the case of $\xi$
Phe, a particularly challenging EB with a 2.6~M$_\odot$ secondary and a
3.9~M$_\odot$ primary that is leaving the main sequence. An {\it ad hoc}
decrease in the metallicity of the secondary is required to improve the fit,
but even without such an adjustment the fit to both components is marginally
acceptable ($\chi_\nu^2 \approx 4$) and has $\Delta\tau\approx 3$\%.

\subsection{Solar-mass stars ($M \approx 1$ M$_\odot$)}

\begin{figure}[h]
% \vspace*{-2.0 cm}
\begin{center}
 \includegraphics[width=2.2in]{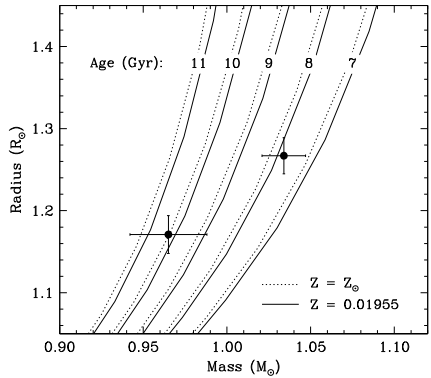} 
% \vspace*{-1.0 cm}
 \caption{Theoretical evolutionary model fits to the components of the
EB CV Boo by \cite[Torres et al.\ (2008)]{torres08}. The active secondary 
of this solar-mass system is 10\% larger than predicted by the 8 Gyr
model isochrone, leading to a
large age discrepancy between the components of $\Delta\tau\approx 25$\%.
 }
   \label{CVBoo}
\end{center}
\end{figure}

At masses of approximately 1 M$_\odot$, the theoretical models begin to
show larger systematic discrepancies in the predicted ages. A good example
is CV Boo, an old main-sequence EB with component masses of 1.03 and 0.97
M$_\odot$ (\cite[Torres et al.\ 2008]{torres08}). The primary shows evidence
for having entered the H shell-burning stage, for which the predicted
model age of 8 Gyr is in good agreement (Fig.\ \ref{CVBoo}). However,
the secondary appears to be 25\% older due to its radius being $\sim 10$\%
larger than predicted by the 8 Gyr isochrone. The oversized radius of the
secondary is likely due to its magnetic activity (see \S\ref{activity}).

Of course, the Sun is the only star for which an absolute age can be
determined directly (e.g., chemical dating of meteorites). While the Sun's
physical properties can by matched to better than 1\% at the solar age by models
that incorporate all of the observational constraints (including, e.g.,
helioseismology), the Sun's age cannot be predicted to better than $\sim
7$\% if given only its observed mass, radius, temperature, and metallicity
(\cite[Young \& Arnett 2005]{young05}). This is likely to be the best
absolute accuracy that can be achieved with current models applied to
EBs with a similar set of observational constraints.

%Another solar-mass exemplar is the pre--main-sequence EB V1174 Ori

\subsection{Low-mass stars ($M < 0.8$ M$_\odot$)}

\begin{figure}[bh]
\begin{center}
 \includegraphics[width=2.4in]{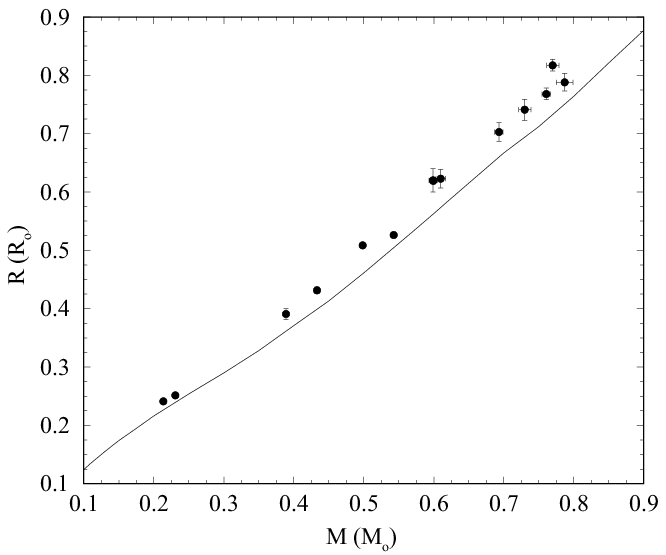} 
 \includegraphics[width=2.35in]{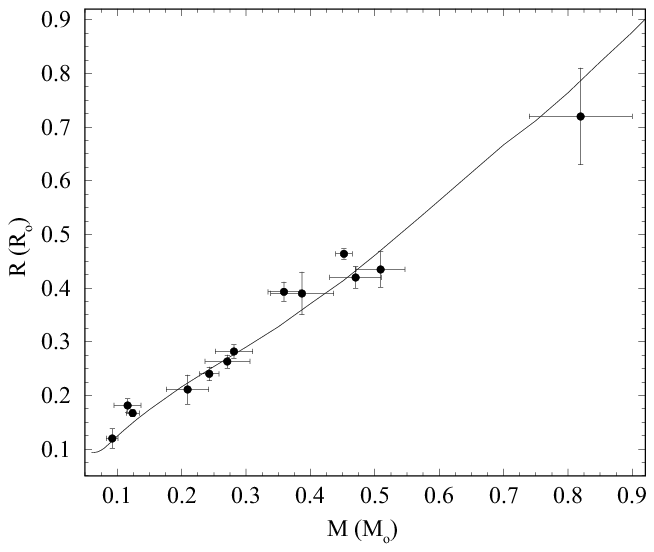} 
 \caption{{\it Left:} Compilation of low-mass EB measurements, showing that the
observed radii of these active stars are systematically larger by 10--20\% 
than predicted.
The solid line is a 1 Gyr isochrone from the models of \cite[Baraffe et al.\
(1998)]{baraffe98}. {\it Right:} Same, but for single-lined EBs,
which are effectively single stars from the standpoint of tidal effects which
may induce activity. These inactive stars' radii agree much better 
with predictions. Note that the masses of single-lined EBs are model dependent
and hence less accurate.  Adapted from \cite[Ribas et al.\ (2008)]{ribas08}.
 }
   \label{bigradii}
\end{center}
\end{figure}

The past several years have seen rapid progress in the number of low-mass
EBs that have been discovered and their components analyzed. A consistent
finding among these studies is that the observed stellar radii are 10--20\%
larger than predicted by the models. 
For example, \cite[L\'{o}pez-Morales (2007)]{lopez07} and \cite[Ribas
et al.\ (2008)]{ribas08} have compiled the literature data for low-mass EBs
with $0.2 < M/{\rm M}_\odot < 0.8$. They find that in virtually all cases
the theoretical main sequence predicts radii smaller than those observed
(Fig.\ \ref{bigradii}). 
These oversized radii make the stars appear 50--90\% older or younger 
than expected (depending on whether post-- or pre--main-sequence
models are used; see also Fig.\ \ref{CVBoo}). 

\begin{figure}[th]
\begin{center}
 \includegraphics[width=2.6in]{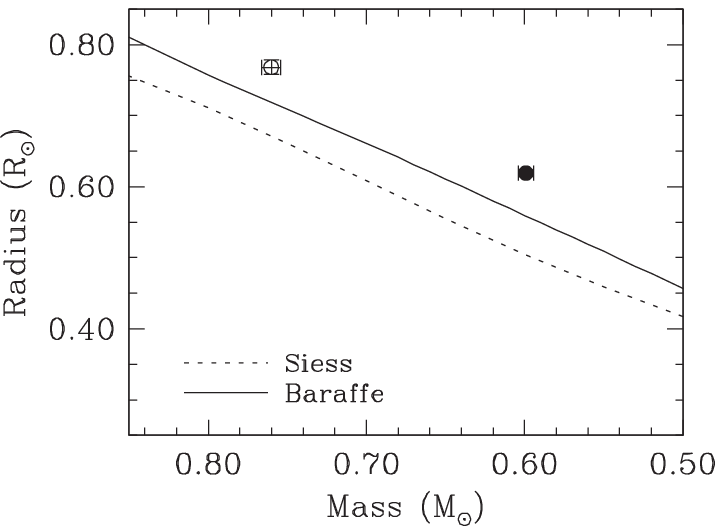} 
 \includegraphics[width=2.3in]{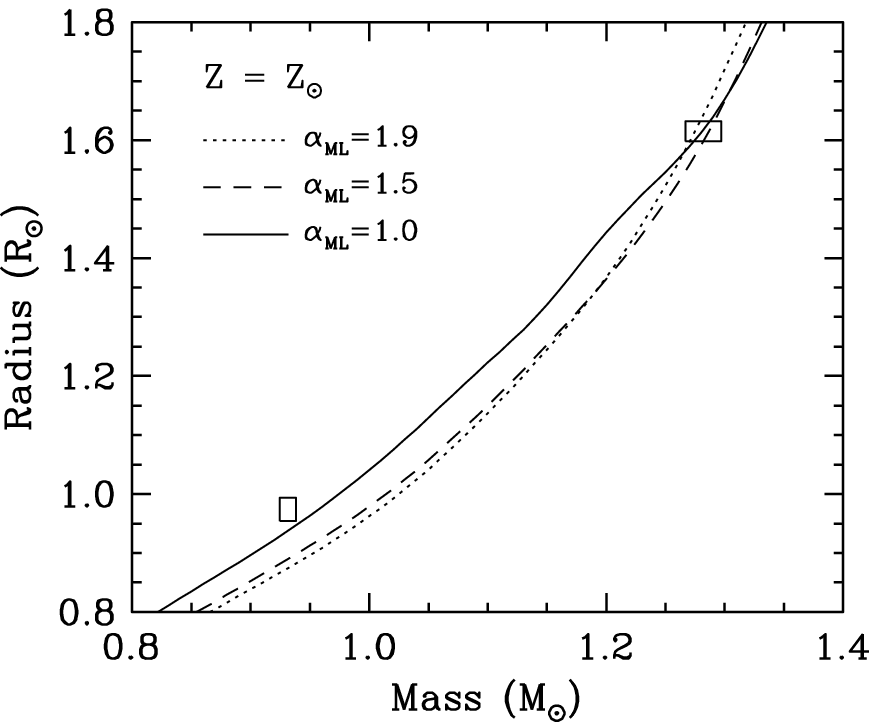} 
 \caption{{\it Left:} Oversized radii are confirmed for the active, low-mass
components of the EBs YY Gem (filled) and V818 Tau (open), for which 
independent age estimates have been made from their membership in
young comoving groups.
Adapted from \cite[Torres \& Ribas (2002)]{torres02}.
{\it Right:} The components of V1061 Cyg are compared with isochrones from
\cite[Baraffe et al.\ (1998)]{baraffe98} with different values of the 
convective mixing length, $\alpha$. The oversized radius of the low-mass secondary
requires suppressed convection (small value of $\alpha$). 
Adapted from \cite[Torres et al.\ (2006)]{torres06}.
 }
   \label{YYGem}
\end{center}
\end{figure}

Importantly, there are now several low-mass EBs for which there exist
independent age constraints (e.g., YY Gem, V818 Tau, V1061 Cyg), and in
these systems the same age discrepancies are verified (Fig.\ \ref{YYGem}).
A few EBs in young open clusters have also been found
(e.g.\ \cite[Hebb et al.\ 2006]{hebb06}; 
\cite[Southworth et al.\ 2004]{southworth04}), again verifying these trends.
More EBs such as these with independent age determinations are very much needed.

\section{The effect of activity in low-mass stars\label{activity}}

There is now very good evidence that the unexpectedly large radii of
low-mass EBs is related directly or indirectly to magnetic activity on
these stars. Several of the authors who published the original analyses 
of low-mass EBs had noted that the stars showing
larger-than-predicted radii also show evidence for activity, in the form of
H$\alpha$ emission, X-rays, spot-modulated light curves, and other tracers.

\begin{figure}[bh]
\begin{center}
 \includegraphics[width=3.85in]{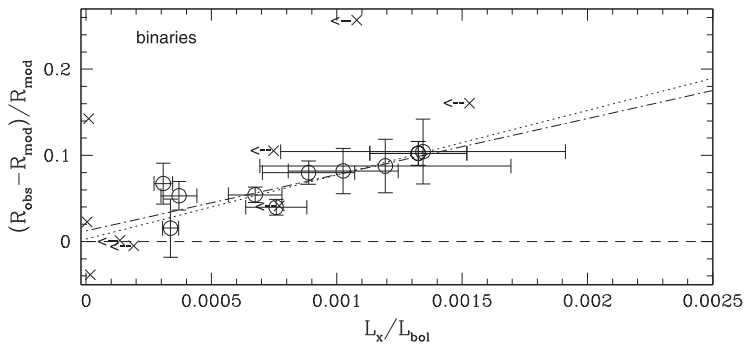} 
 \caption{Correlation between X-ray luminosity and fractional discrepancy
between measured and predicted radii for low-mass eclipsing binaries.
Adapted from \cite[L\'{o}pez-Morales (2007)]{lopez07}.
 }
   \label{activity-fig}
\end{center}
\end{figure}

More recently, \cite[L\'{o}pez-Morales (2007)]{lopez07} has
demonstrated the relationship explicitly (Fig.\ \ref{activity-fig}). This
is very good news, not only because it points clearly to an underlying cause
for the observed oversized radii, but also because the tight correlation
with X-ray luminosity suggests that this effect can be calibrated and the
ages corrected. Indeed, single-lined EBs---which can be regarded
as effectively single stars and which are thus less likely to have magnetic
activity driven through interactions with a companion---do not show 
systematically oversized radii (Fig.~\ref{bigradii}, right).

\begin{figure}[th]
\begin{center}
 \includegraphics[width=5.3in]{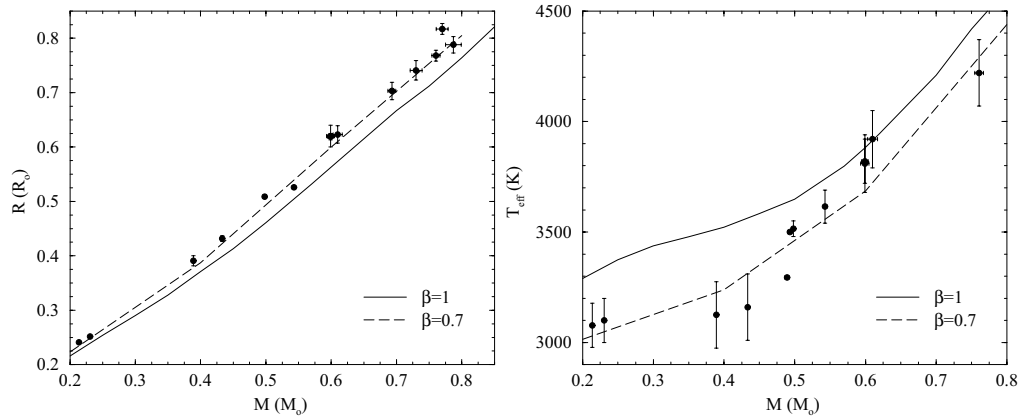} 
 \caption{Model isochrones (\cite[Chabrier et al.\ 2007]{chabrier07})
incorporating activity in low-mass stars. 
{\it Left:} Oversized radii of active, low-mass EBs are
well fit by a 1 Gyr isochrone from \cite[Baraffe et al.\
(1998)]{baraffe98} adopting a spot-covering fraction of 0.3 (i.e.\ 70\%
percent of the stellar surface is free of spots). {\it Right:} Same, but 
for effective temperature. Adapted from \cite[Ribas et al.\ (2008)]{ribas08}.
 }
   \label{chabrier-fig}
\end{center}
\end{figure}

In addition, recent modeling that incorporates the effects of magnetically
suppressed convection in low-mass stars due to magnetically active surfaces
is now able to fit the observed oversized radii of active EBs extremely
well (Fig.\ \ref{chabrier-fig}). In addition, these models simultaneously
can explain the systematically low effective temperatures of these stars.

In these new models, strong magnetic fields cause a suppression
of convection near the surface. Heat flow to the surface is inhibited 
(by analogy to dark sunspots on the Sun), resulting in a decrease in the
star's effective temperature. However, the star's overall luminosity is
roughly fixed by internal processes far removed from the surface boundary
condition, and thus the star's radius adjusts to a larger size in order
to radiate the flux. 

It should be stressed that at present these models use
parametrizations of surface spots and of suppressed convection in place of
a full physical treatment of convection and surface fields.
Even so, several additional lines of evidence corroborate this general picture.
First, the observed properties of young, low-mass EBs are in general best fit
by model isochrones with low convective efficiency, $\alpha \sim 1$
(e.g.\ \cite[Mathieu et al.\ 2007]{mathieu07}). Second, the observed surface 
lithium abundances of young, low-mass EBs clearly indicate weak convective
mixing (e.g.\ \cite[Stassun et al.\ 2004]{stassun04}). Third, analyses of
low-mass EBs have found that indeed the luminosities of the stars are in
good agreement with the models even when the radii and temperatures are
very discrepant. For example, in the brown-dwarf EB 2M0535--05
(\cite[Stassun et al.\ 2006]{stassun06}), the 
brown dwarfs display $\sim 10\%$ oversized radii, and the temperature of
the very active primary (\cite[Reiners et al.\ 2007]{reiners07})
has been so severely suppressed that it is in fact
cooler than the lower-mass secondary. However, the luminosities 
remain in good agreement with model predictions for brown dwarfs at an
age of $\sim 1$ Myr (\cite[Stassun et al.\ 2007]{stassun07}). 

Finally, these findings have implications for low-mass
stars more generally. First, because activity has the effect of decreasing
the effective temperature but leaving the luminosity relatively unaffected,
we can expect to see these stars scattered horizontally in the H-R diagram.
Second, these effects will need to be taken into account when deriving
ages from other means, such as age-activity relations and surface lithium
abundances.

\section{Star formation effects at very young ages}

Testing the accuracy of stellar evolutionary models via the $\Delta\tau$
test, as we have done above, assumes that EBs
represent {\it coeval} systems of two stars that formed from the
same material at the same time. Indeed, in many cases, this assumption
of the coevality of EB components has been used to calibrate the various
input parameters of the evolutionary tracks. For example, 
\cite[Young \& Arnett (2005)]{young05} have adjusted model parameters
such as core overshooting, and have determined secondary stellar properties
such as metallicities, on the basis of requiring that the evolutionary
tracks yield the same model ages for the two stars of an EB. Similarly,
\cite[Luhman (1999)]{luhman99} has adjusted the temperature scale of
young, low-mass stars on the basis of requiring that pre--main-sequence
evolutionary tracks yield coeval ages for the components of pre--main-sequence
binaries.

\begin{figure}[t]
\begin{center}
 \includegraphics[width=5.3in]{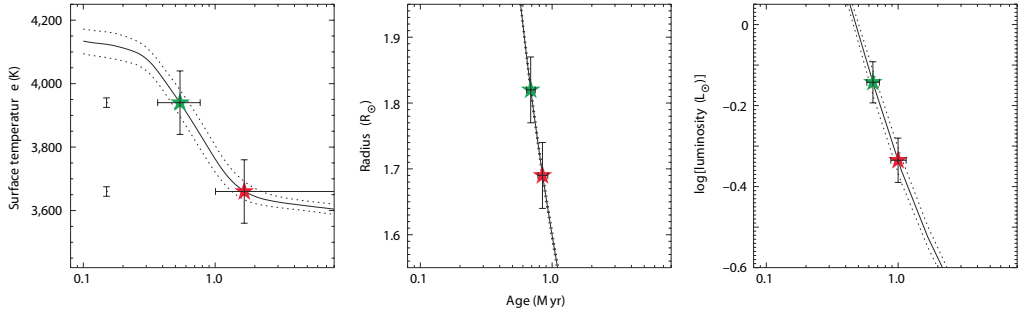} 
 \caption{Comparison of physical properties of Par 1802 with theoretical
predictions. In each panel,
the solid line shows the predicted evolution of a 0.41 M$_\odot$ star 
from the models of 
\cite[D'Antona \& Mazzitelli (1997)]{dantona97}. Dotted lines show the
result of changing the stellar mass by 0.015 M$_\odot$
uncertainty in the measured masses. Vertical error bars on the 
points represent the combination of measurement and systematic uncertainties.
Horizontal error bars represent the range of ages for which the
theoretical models are consistent with the measurements within the
uncertainties (including systematic uncertainties). 
Note that the uncertainties in the temperatures, radii and luminosities are
not independent between the two stars, because they are connected by
precisely determined ratios; thus, for example, the primary star cannot be
forced cooler while simultaneously forcing the secondary warmer. The nominal 
age of the Orion nebula cluster in which this EB is found is $\sim 1$ Myr.
Adapted from \cite[Stassun et al.\ (2008)]{stassun08}.
 }
   \label{stassun}
\end{center}
\end{figure}

We now have evidence that, in at least some cases, the components of
very young binaries may not in fact be strictly coeval. In particular,
Par 1802 is a recently discovered EB in the Orion Nebula, with a mean
age of $\sim 1$ Myr, whose components are identical in mass to within 2\%
($M_1 = M_2 = 0.41$ M$_\odot$; \cite[Stassun et al.\ 2008]{stassun08}).
Having the same mass, these `identical twin' stars are predicted by the
models to have identical temperatures, radii, and luminosities. However,
the components of Par 1802 are found to have different temperatures 
($\Delta T \approx 300$K, or about 10\%), radii that differ by 5\%, and
luminosities that differ by a factor of $\sim 2$ (Fig.\ \ref{stassun}).
These surprising dissimilarities between the two stars can be interpreted
as a difference in age of $\Delta\tau \sim 30$\%. It has been speculated
that this age difference likely reflects differences in the star formation
history of the two stars, differences that may be specific to 
binary star formation \cite[Simon et al.\ (2009)]{simon09}.

Unfortunately, if such non-coevality turns out 
to be a common feature of young binaries, then Par 1802 suggests
that using very young EBs to calibrate the evolutionary model ages may 
be limited to $\sim 30$\% accuracy. Fortunately, these effects
are largely erased after $\sim 10$ Myr. For example, 
\cite[Stempels et al.\ (2007)]{stempels07} find that the components
of ASAS J052821$+$0338.5, a solar-mass EB with an age of 12 Myr, are
coeval to $\Delta\tau \sim 10$\%.

\section{The future of eclipsing binaries with large surveys}

The central importance of EBs for stellar age determinations
implies an ongoing need for precise and accurate EB data. As
sky surveys are gaining on both precision and diversity, and since more
and more medium size observatories are being refurbrished into fully
robotic telescopes, there is a ``fire-hose" of photometric and
spectroscopic data coming our way. Methods to reduce and analyze the data
thus cannot rely on manual labor any longer; rather, automatic approaches
must be devised to face the challenge of sheer data quantity. Pioneering
efforts of automating the analysis of survey data by several groups,
most notably Wyithe \& Wilson (2001, 2002), Wyrzykowski et al.\ (2003),
Devor (2005), and Tamuz et al.\ (2006). These are reviewed in 
Pr\v sa \& Zwitter (2007). 

\begin{figure}[th]
\begin{center}
 \includegraphics[angle=270,width=5.3in]{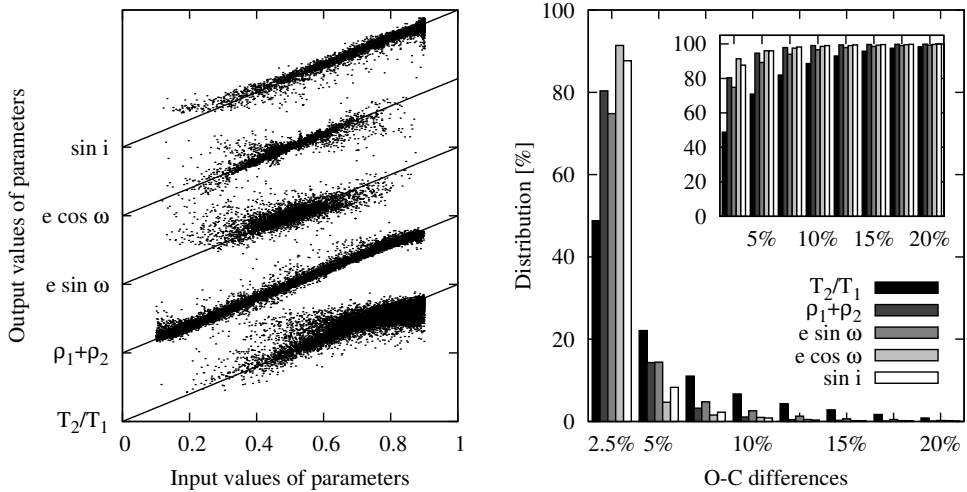} 
 \caption{ Neural network recognition performance on 10,000 detached
EB light curves. {\it Left:} comparison between input and output values of
parameters. {\it Right:} distribution of differences (main panel) and their
cumulative distribution (inset). 
Adapted from \cite[Pr\v{s}a et al.\ (2008)]{prsa08}.
 }
   \label{prsa}
\end{center}
\end{figure}

A recent stab at automation is implemented within the Eclipsing
Binaries via Artificial Intelligence project (EBAI; Pr\v sa et al.\ 2008). A
back-propagating neural network is applied as a non-linear regression tool
that maps EB light curves onto a subset of parameter space that is sensitive
to photometric data. Its performance has been thoroughly tested on detached
EB light curves (Fig.\ \ref{prsa}) and applied successfully to OGLE data. In a
matter of seconds, the network is able to provide principal parameters of
tens of thousands of EBs. The results that come from such an engine may be
readily used to select those EBs that are most interesting for the studies
of stellar formation and evolution. Given the number of surveys, we are
talking thousands of interesting EBs! Since our understanding relies on
these systems, such a disproportionate jump in data quantity will surely
provide further insights and enhance statistical significance of our results.

%\begin{discussion}

%\end{discussion}

\end{document}